# *Stress, Depression and Neuroplasticity*

*Shatrunjai P. Singh[1] and Swagata Karkare[2]*

[1]Lindner College of Business, University of Cincinnati, Ohio, USA

[2]School of Public Health, Boston University, Boston, MA, USA

## Abstract

Modifications of signaling pathways and synapses owing to changing behaviors, environments, numerous neural modulation as well as brain-tissue injuries is defined as neuroplasticity in developmental neurology. The central purpose of the review is to gain a better understanding of the relation between stress, depression and neuroplasticity and explore potential therapeutic interventions for enhancing neural resilience. We have also reviewed the role of different factors like age, stress and sex on inducing neuroplasticity within various brain regions.

## 1. Introduction

Modifications of signaling pathways and synapses owing to changing behaviors, environments, various neural processes as well as brain-tissue injuries is defined as neuroplasticity in developmental neurology (Pascual-Leone et al., 2011). Neuroplasticity aids the brain in processing sensory inputs as well as creating suitable adaptive responses to consequent stimuli. Nobel laureate Eric Kandel once said, "Neuroplasticity is what endows each of us with our individuality". This review focuses mainly on how stress can divert the protective influence of neuroplasticity to instead become harmful. A profound understanding of this will help to promote a mechanism of 'resilience' present in the diseased brain using anti-depressant drugs as a novel approach to treat stress-related brain disorders as well as related mood disorders, specifically depression and anxiety. The central purpose of the review is to gain a better understanding of the relation between stress, depression and neuroplasticity and explore potential therapeutic interventions for enhancing neural resilience.

Brain circuitry can be remodeled by experience (Bennett et al.,1964), and stressful experiences have functionally relevant effects on dendritic arbor, spine, and synapse number in many brain regions, including the hippocampus, amygdala, and the prefrontal cortex (PFC), with effects not only on cognitive function but also on emotional regulation and other self-regulatory behaviors (McEwen and Gianaros, 2011). Stress can have a profound effect on the PFC in particular. The PFC is important for working memory i.e. the ability to keep events in mind and perform self-regulatory and goal-directed behaviors. Structural and functional plasticity in this brain region illustrates the profound capacity of behavioral experiences to change neural circuitry and alter brain function, with the most significant impact occurring during early childhood and adolescence.

## 2. Role of stress in neuroplasticity

Many factors are known to impact neuroplasticity and cellular resilience. These include alterations in the Hypothalamic-Pituitary-Adrenal (HPA) axis and glutamate neurotransmission as well as impaired neurotrophic/neuroprotective signaling. Further, stress also affects neuronal morphology. For instance, repeated restraint stress can lead to atrophy and death of CA3 pyramidal neurons in the hippocampus in both rodents and non-human primates (Sapolsky 1996; McEwen 1999). The site specificity of these gross morphological changes due to stress is noteworthy. Some significant instances of these gross morphological changes include dendritic shortening in the medial prefrontal cortex (Cerqueira, 2007; Cook & Wellman, 2004; Liston et al., 2006; Radley, Sisti & McEwen, 2004), but dendritic growth of neurons in basolateral amygdala ((Vyas, Mitra, Rao & Chattarji, 2002)), as well as in orbitofrontal cortex (Liston et al., 2006) after subjecting experimental models to chronic immobilization stress.

Atrophy is equivalent to a decrease in the number and length of branch points of the apical dendrites of CA3 neurons. Dentate granule cells in the hippocampus appear comparatively resistant to atrophy and death. Neurogenesis is the mitosis and generation of progenitor cells in regions of the adult mammalian brain such as the dentate gyrus of the hippocampus which ultimately differentiate into functionally integrated neurons throughout life. However, stress has shown to reduce the neurogenesis of these cells in adult animals (Gould & Cameron, 1997, 1998). It is believed that neurogenesis provides an 'enriched' environment that greatly contributes to cognitive processes like memory and learning (Kempermann, Kuhn & Gage, 1997; Van Praag, Kempermann & Gage, 1999). Brain-derived neurotrophic factor (BDNF), a trophic involved in survival of striatal neurons in the brain, regulation of stress response and in the biology of mood disorders, is induced in response to neuronal activity, and has been shown to play a critical role in cellular models of learning and memory (i.e., long-term potentiation or LTP).Decreased BDNF expression in the CA1 and CA3 pyramidal and dentate granule cell layers of hippocampus after acute or repeated immobilization stress has been shown by researchers (Smith, Makino, Kvetnanský & Post, 1995).

## 3. Role of age and gender in neuroplasticity

Age is also an important factor in neuroplasticity. Studies have reported behavior induced neuronal shrinkage and resilience, predominantly in the distal apical dendrites of young adults. Importantly, this capacity was lost in adult or middle aged rats (Bloss, Janssen, McEwen & Morrison, 2010). It is now known that effects of chronic stress carry over to older ages; in adult rats, 21 days of chronic restraint stress impaired working memory and caused spine loss and debranching of dendrites on the medial PFC neurons (Hains et al., 2009). It is known that in addition to aging, there are also sex differences in responses of neuroplasticity to stress. Specifically, in males, pyramidal neurons in the layer III are affected along with apical dendritic length shrinkage. This shrinkage is most prominent in the distal apical dendritic branches amongst the thin spines and it is often accompanied by spine loss of approximately 30% of axospinous synapses (Bloss et al., 2010, 2011; Cook and Wellman, 2004; Radley et al., 2004). Taken together, evidence showing that the mature brain has greater capacity for plasticity than previously believed is carving a path for future behavioral and pharmacological-based therapies that harness neural plasticity for recovery.

While there can be dramatic morphologic changes, they may not be permanent. In the absence of stress, neurons, especially in young animals, recover structurally and functionally almost completely within 3 weeks (Bloss et al., 2011; Radley et al., 2005). Although spine density is only partially recovered, the dendritic arbor recovers fully and this has important implications for therapeutic interventions because spine loss impairs cognitive processes, especially working memory performances (Hains et al., 2009).

Various parts of the brain interconnect with each other, either directly or indirectly. The prefrontal cortex, amygdala and hippocampus interconnect and influence each other via direct as well as indirect neural activity. Again, this is another important aspect which can be harnessed while devising therapeutic interventions. Of special interest are findings that amygdal inactivation blocks stress induced hippocampal long term potentiation and spatial memory (Kim et al., 2005) while stimulation of basolateral amygdala enhances dentate gyrus field potentials (Ikegaya et al., 1996). Another study also reported decreased responsiveness of central amygdal output neurons upon stimulation of the medial prefrontal cortex (Quirk et al., 2003). Thus, the amygdala and hippocampus act in conjunction. Inhibition or stimulation of either one has a directly proportional effect on the other and hence this connection forms the basis of important therapeutic targets.

## 4. Potential therapeutic targets

McEwen and colleagues examined the effects of antidepressant treatment on the atrophy of CA3 pyramidal neurons and demonstrated that an atypical drug-Tianeptine (Stablon) which is a Selective serotonin reuptake enhancer (in contrast to most antidepressant agents) and not a typical 5-HT selective reuptake inhibitor (like fluoxetine) was more effective in enhancing neural resilience (Watanabe et al 1992). Antidepressant treatment is believed to act in a number of ways to counteract the effects of chronic stress. These include upregulation of the neurogenesis of dentate gyrus granule neurons and BDNF, especially in the hippocampus and by blocking downregulation of BDNF in response to stress. BDNF is a good potential target because studies have shown that direct application of BDNF into the midbrain of rats is reported to have antidepressant effects in behavioral models of depression, including the forced swim and learned helplessness paradigms (Siuciak et al 1996). BDNF is also a potent neurotrophic factor for both the NE and 5-HT neurotransmitter systems. In addition to these, modulation of the altered HPA axis or glutamergic activity or the signaling cascades, possibly via direct circuit stimulations using techniques like deep brain stimulation, magnetic stimulation or vagus nerve stimulation could make promising experimental approaches.

Therapeutic interventions that could change neural architecture and improve cognition would be beneficial in stress-induced dysfunction. Some studies have demonstrated the ability of estrogen to potentiate stress-induced plasticity (Shansky, Rubinow, Brennan & Arnsten, 2006). Studies of connectivity between the prefrontal cortex, amygdala, and hippocampus to elucidate that their functional relationships may accelerate the development of such therapies. Understanding the molecular and cellular mechanisms of neuroplasticity, including but not limited to signal transduction and gene expression, structural alterations of neuronal spines and processes, and neurogenesis will lead to novel drug targets that could prove to be effective and rapidly acting therapeutic interventions. In addition, studies related to the neuroplastic responses to various disorders like depression and anxiety are of high significance in the fast advancing field of neuroscience. Neuronal resilience or the ability of neurons to reverse the alterations (in terms of the structure as well as function) makes this aspect a very promising treatment for a number of mental disorders (Hester et al., 2016; Karkare et al., 2014; Singh, 2015; Singh et al., 2008, 2013, 2015, 2016).

## 5. Conclusion

In conclusion, it is critical to primarily gain a better understanding of the contribution of early life experiences like adverse life-events on plasticity. Since they are believed to have effects on neuroplasticity, it would be worthwhile to explore if researchers can take advantage of neuro-resilience seen in young rodent models and develop therapeutic interventions to undo the effects of stress by enhancing neuro-resilience alongside neuroplasticity. The next step would be to evaluate if there's a way to retain resilience and plasticity of prefrontal neurons as humans age. The main aim of this review was to focus on the effects of stress on prefrontal cortical plasticity. Pioneering work on reorganization of the adult cerebral cortex and the reversal of developmentally induced monocular deprivation in visual cortex provides further impetus to some likely therapeutic strategies. Hence, interventions that can change brain architecture and help improve cognitive function and self-regulatory behaviors certainly hold tremendous potential. According to Bavelier et al., ongoing studies at the cellular and molecular level are beginning to reveal mechanisms involving perineuronal nets and excitatory/inhibitory balance as possible intervention strategies (Bavelier et al., 2010). Additionally, one can also look at possibilities of non-pharmacological interventions like yoga, acupuncture, exercise, meditation as adjuvants to anti-depressant drug therapies and whether pharmacological and non-pharmacological treatments together can give more beneficial results in terms of increased neurogenesis instead of either one of them alone.

Finally, advancing techniques like optogenetics combined with modern imaging methods can greatly accelerate our understanding of neuroplasticity and vulnerability of the various brain regions, especially the prefrontal cortex, spanning the entire life course of human beings. These studies can also add to the knowledge of homologous region adaptations. For instance, if brain damage affects one side of the parietal lobe, then can the other side reorganize itself to replicate the various forms of information previously stored in the affected side? This knowledge will eventually be helpful for developing therapeutic interventions that promote mental and cognitive health by enhancing synaptic properties and neural circuit characteristics.